\documentclass[hidelinks,10pt,aps,prc,twocolumn,showpacs,showkeys,amsmath,floatfix,superscriptaddress,preprintnumbers]{revtex4-1}
\usepackage[utf8]{inputenc}
\usepackage{color,graphicx}
\usepackage{dcolumn}                 % Align table columns on decimal point
\usepackage{bm}                      % bold math
\usepackage{ulem,soul}
\usepackage{hyperref}
\usepackage{tabularx,booktabs}
\usepackage{amssymb}
\usepackage{amsmath}
\usepackage{bm}
\usepackage{nicefrac}
\usepackage{multirow}

\DeclareMathOperator{\tr}{Tr}

\usepackage[dvipsnames,usenames]{xcolor}
\DeclareMathAlphabet{\mathpzc}{OT1}{pzc}{m}{it}

\hypersetup{
  colorlinks   = true, %Colours links instead of ugly boxes
  urlcolor     = blue, %Colour for external hyperlinks
  linkcolor    = blue, %Colour of internal links
  citecolor   = red %Colour of citations
}

\begin{document}

%\preprint{INT-PUB}

\title{Quantifying the uncertainties on spinodal instability in stellar matter through meta-modeling}

\author{Sofija Anti\'{c}}
\affiliation{CSSM, Department of Physics, University of Adelaide, SA 5005, Australia}

\author{Debarati Chatterjee}
\author{Thomas Carreau}
\author{Francesca Gulminelli}
\affiliation{CNRS, ENSICAEN, UMR6534, LPC, F-14050 Caen cedex, France}

\date{\today}

\begin{abstract}

%Correction by Debi:

The influence of the uncertainties of the equation of state empirical parameters on the 
neutron stars crust-core phase transition  is explored within a meta-modeling 
approach, in which the energy per particle is expanded as a Taylor series in density and
asymmetry around the saturation point. The phase transition point is estimated from the 
intersection of the spinodal instability region for dynamical fluctuations with the 
chemical equilibrium curve. Special attention is paid to the inclusion of high-order 
parameters of the Taylor series and their influence on the transition point.
% The influence of the equation of state empirical parameters uncertainties on the crust-core
% phase transition point for neutron stars is explored within a meta-modeling approach. The 
% transition point is estimated using the dynamical spinodal method, and special attention 
% is paid to the inclusion of high-order parameters and their influence on the transition point.
An uncorrelated prior distribution is considered for the empirical parameters, with bulk 
properties constrained through effective field theory predictions, while the surface parameters
are controlled from a fit of nuclear masses using the extended Thomas Fermi approximation.
The results show that the isovector compressibility $K_{sym}$ and skewness $Q_{sym}$ have the most significant correlations 
with the transition point, along with the previously observed influence of the 
$L_{sym}$ parameter. The estimated density and pressure of the crust-core transition are 
$n_t = (0.071 \pm 0.011)$ $fm^{-3}$ and $P_t = (0.294 \pm 0.102)$ $MeV fm^{-3}$.
\end{abstract}

\maketitle

\section{Introduction}
%----------------------------------------------------------------------------------------

 The study of nuclear matter under different conditions has been an inexhaustible source of
 both experimental and theoretical research for a long time. The properties of such matter 
 change profoundly with the increase of density $n_0$ and/or temperature $T$, and a complex 
 phase diagram is predicted, see for a recent review Ref. \cite{Oertel:2016bki}.
 
 For the case of $T=0$, most of the nuclear matter features are known through the study of 
 nuclei that span the valley of stability, having densities below and around saturation 
 $n_{sat} \sim 0.16$ $fm^{-3}$. Additionally, the observable neutron stars (NSs) are a very 
 rich source of information over a higher range of nuclear densities. While there is no
 consensus yet on whether the NS core contains hadron matter only or there are exotic 
 particles coming into play \cite{Oertel:2016bki}, %(since the densities could be over $10$ $n_{sat}$)
 there is a general agreement that the nuclear matter goes through a continuous phase
 transition from the homogeneous liquid core phase to the inhomogeneous solid NS crust 
 layer of densities close to $n_{sat}$, at around $1$ $km$ from the NS surface. 
 %The transition point is defined as the density at which the homogeneous nuclear matter becomes 
 %unstable with  respect to the density fluctuations. 
 
 The precise density %4 Debi: and pressure 
 at which this happens changes the NS properties, such as for 
 instance the NS radius, expected to be measured up to $5\%$ precision via the Neutron 
 star Interior Composition Explorer (NICER) \cite{RevModPhys.88.021001, Gendreau2012, Bogdanov2012}, and the 
 crustal fraction of the moment of inertia, which is believed to be linked to the observed pulsar
 glitches ~\cite{Haskell2015,Link1999,Steiner2015}.
% installed  to International Space Station (ISS) in June 2017. 
 
 The search for the transition point can be described in a simplified way if we assume that 
 nuclear matter is homogeneous and uncharged. In that case, the transition is known to correspond
 to a first order liquid-gas (LG) phase transition which can be tracked by the thermodynamical 
 spinodal, defined as the density locus at which homogeneous matter becomes unstable with 
 respect to density fluctuations. In reality, the situation is more complicated since the NS crust
 is composed of clusterized matter where, in addition, both surface and Coulomb energy contribute 
 to the equilibrium energy, as discussed in section \ref{sec:MM_th}. The additional inclusion of
 finite-size density fluctuations will define the so called dynamical spinodal and give a better 
 estimation of the transition density and pressure, which define the crust-core (CC) transition 
 point \cite{PETHICK1995675}.
 
 Different models, both relativistic \cite{Horowitz2001, PhysRevC.74.035802, Moustakidis:2010zx, 
 Fattoyev:2010tb, Cai:2011zn, Pais:2016nzh} and non-relativistic \cite{Vidana:2009is,
 Li:2015qci, Ducoin:2011fy, Routray:2016yvp, Gonzalez-Boquera:2017uep}, give different
 predictions for the transition point. These differences come from many sources 
 (different approaches, different equations of state (EoS), parameter space etc), and 
 it is not easy to make a direct comparison among them and sort out the dependence on 
 the nuclear properties, that could be better constrained by dedicated nuclear experiments. 
 In this work we will quantify the uncertainties on the spinodal instability by using a 
 meta-modeling (MM) approach developed in Ref. \cite{Margueron1} and introduced in section
 \ref{sec:MM_th}, which was already successfully applied to describe NSs
 \cite{Margueron2} and finite nuclei \cite{Chatterjee2017}. Using this technique, a great
 number of EoS models can be generated without imposing a-priori correlations among the 
 empirical parameters, which will enable us to extract the parameters that have the strongest 
 influence the transition point. This set of models will be filtered through constraints
 coming both from ab-initio modelling and empirical information from the nuclear masses. 
 The posterior model set, reduced by the application of these filters, thus represents our
 model-independent present knowledge of the nuclear EoS, and will be used to extract the 
 physical correlations and uncertainty estimations.

 The paper is organized as follows: In the first part of  section \ref{sec:MM_th}, the 
 meta-modeling framework, developed in Ref. \cite{Margueron1} to describe nuclear matter and 
 used in this work, is briefly summarized. The second part of the same section discusses the
 spinodal calculation, necessary to obtain the crust-core phase transition point.
 Section \ref{sec:MMspinodal} elaborates on the importance of a correct spinodal reproduction
 of reference models within the MM approach and the possible sources of deviation. The sensitivity
 study of the LG (CC) transition point with the empirical EoS parameters is summarized in 
 section \ref{sec:Sensitivity}. The Bayesian analysis of the large set of MMs and their 
 parameters is presented in section \ref{sec:Bayesian}, where the importance of the 
 higher-order (HO) parameters in the MM approach is emphasized in the search of the 
 phase transition point to the NS crust. The final section \ref{sec:Conclusion} lists 
 the conclusions of the study.

\section{Theoretical framework for describing nuclear matter}\label{sec:MM_th}

\subsection{Meta-modeling (MM) approach to the EoS}

The meta-modeling approach developed in Ref. \cite{Margueron1} is a theoretical 
framework inspired by a Taylor expansion around the 
saturation density $n_{sat}$  of symmetric nuclear matter, parameterized in terms of the empirical parameters. 
In this section, we briefly recall the main features of the nuclear EoS meta-modeling
method used in this work, for more details see Ref. \cite{Margueron1}.

Nuclear matter, composed of protons and neutrons, is characterized by the isoscalar 
(IS) $n_0 = n_n + n_p$ and isovector (IV) $n_1 = n_n - n_p$ densities where neutron 
and proton densities, $n_n$ and $n_p$, are expressed in terms of the Fermi momentum 
$k_{F_{n,p}}$ as 
%\begin{equation}
$ n_{n,p} = k^3_{F_{n,p}}/(3\pi^2)$.
%\end{equation}
By introducing the asymmetry parameter $\delta=n_1/n_0$, the energy per particle of 
asymmetric matter (ANM) can be separated into the IS and IV channels 
\cite{Dutra2012, Ducoin2010}, as
\begin{equation}\label{eq:en_particle}
 e(n_0, n_1) = e_{IS}(n_0) + \delta^2 e_{IV}(n_0),
\end{equation}
where both can be expanded as series in parameter $x = (n_0 - n_{sat}) / 3 n_{sat}$ 
\cite{Piekarewicz2009}
 \begin{align}\label{eq:expansion}
  e_{IS} = & E_{sat} + \frac12 K_{sat} x^2 + \frac{1}{3!} Q_{sat} x^3 + 
  \frac{1}{4!} Z_{sat} x^4 + ... , \\
  e_{IV} = & E_{sym} + L_{sym} x + \frac12 K_{sym} x^2 \nonumber \\
  & \quad + \frac{1}{3!} Q_{sym} x^3 + \frac{1}{4!} Z_{sym} x^4 + ...
 \end{align}
The coefficients of the expansion define the nuclear empirical parameters which
are ordered according to their power in the density expansion and characterize 
the general properties of relativistic and non-relativistic nuclear interactions.
In the IS channel, the parameters are the saturation energy $E_{sat}$, %the saturation density $n_{sat}$, 
the incompressibility modulus $K_{sat}$, the isoscalar skewness
$Q_{sat}$, and the isoscalar kurtosis $Z_{sat}$; the IV channel defines the symmetry
energy $E_{sym}$, the slope $L_{sym}$, the isovector incompressibility $K_{sym}$, the
isovector skewness $Q_{sym}$, and the isovector kurtosis $Z_{sym}$.

The formalism used in this work is the ELFc meta-modeling introduced in Ref. 
\cite{Margueron1}. The idea is to sort out the known non-interacting part from 
the total functional $e(n_0,n_1)$:
\begin{equation}%\label{eq:en_particle}
 e(n_0, n_1) = e_{FG}(n_0, n_1)+e_{int}(n_0,n_1),
\end{equation}
with
\begin{align}
 e_{FG}(n_0, n_1) =& \frac{t^{FG}_{sat}}{2} \Big( \frac{n_0}{n_{sat}}\Big)^{2/3} \nonumber
 \\ & \Bigl[\frac{m}{m^*_n}(1+\delta)^{5/3} + \frac{m}{m^*_p}(1-\delta)^{5/3}\Bigr],
\end{align}
describing the free non-relativistic Fermi gas (FG). In this expression, 
%$t^{FG}_{sat} = \frac{3\hbar^2}{10m}\Big(\frac{3\pi^2}{2}\Big)^{2/3} n_{sat}^{2/3}$ 
$t^{FG}_{sat} = (3\hbar^2 / 10m)(3\pi^2/2)^{2/3} n_{sat}^{2/3}$ 
is the kinetic energy per nucleon of non-interacting symmetric matter (SM) at 
saturation, and $m$ ($m^*_q(n_0,n_1)$) for $(q = n, p)$ is the bare (effective) nucleon mass.
%we can also describe isospin asymmetric matter (ANM) with the boundaries $\delta = 0$ 
%for to symmetric nuclear matter (SNM) and $\delta = 1$ for pure neutron matter (PNM).
Since the functional form of the interaction $e_{int}$ and of the effective masses
$m_q^{\ast}$ is   largely unknown, a Taylor expansion up to order $N$ with a low density
correction is introduced, according to: 
\begin{equation}
 e_{int}^N(n_0, n_1) = \sum^N_{\alpha\geq0} \frac{1}{\alpha!}
 (\nu^{IS}_{\alpha} + \nu^{IV}_{\alpha} \delta^2 ) 
 x^{\alpha} u^N_{ELFc,\alpha}(x),
\end{equation}
where  the low-density correction function $u^N_{\alpha}(n_0, n_1) = 1-(-3x)^{N+1-\alpha} \exp(-b n_0/n_{sat})$ 
is defined in such a way that the limit $e_{int}^N\rightarrow0$ is satisfied for $n_0 \rightarrow 0$. 
The low density correction parameter $b$ is fixed by requiring that the exponential part 
of the correction function at a given low density value $n_0=n_m$ satisfies the condition:
\begin{equation}\label{eq:b_exp}
 exp( -b\frac{n_m}{n_{sat}} ) = \frac12,
\end{equation}
and is fixed by the choice of density $n_m$, where the Taylor expansion is supposed to 
break down. This choice will be discussed in subsection \ref{subsec:Bparam}.

The momentum dependence of the nuclear interaction introduces the concept 
of effective mass $m^{\ast}$, a modified internal mass of the nucleon due 
to the in-medium nuclear interactions. The Landau effective mass is 
parametrized by the introduction of two additional empirical parameters,
$\kappa_{sat}$ and $\kappa_{sym}$,
\begin{align}
 \kappa_{sat}&=\frac{m}{m^\ast(n_{sat},\delta=0)}\\
 \kappa_{sym}&=\frac12 \Big[\frac{m}{m^{\ast}_n(n_{sat},\delta=1)}-\frac{m}{m^{\ast}_p(n_{sat},\delta=1)} \Big], 
\end{align}
in the following manner:
\begin{equation}
 \frac{m}{m^{\ast}_q} = 1 + (\kappa_{sat} + \tau_3 \kappa_{sym} \delta) \frac{n_0}{n_{sat}}.
\end{equation}

A one-to-one correspondence can be obtained between the parameters of the ELFc MM 
$\nu^{is}_{\alpha}$ and $\nu^{iv}_{\alpha}$, $\kappa_{sat}$, $\kappa_{sym}$ 
and the empirical parameters defined in eq. (\ref{eq:expansion}). 
In the isoscalar sector we have: 
\begin{align}
 \nu^{IS}_{\alpha=0} &= E_{sat} - t^{FG}_{sat} ( 1 + \kappa_{sat}),\\
 \nu^{IS}_{\alpha=1} &= - t^{FG}_{sat} ( 2 + 5\kappa_{sat}),\\
 \nu^{IS}_{\alpha=2} &= K_{sat} - 2 t^{FG}_{sat} (-1 + 5\kappa_{sat}),\\
 \nu^{IS}_{\alpha=3} &= Q_{sat} - 2 t^{FG}_{sat} ( 4 - 5\kappa_{sat}),\\
 \nu^{IS}_{\alpha=4} &= Z_{sat} - 8 t^{FG}_{sat} (-7 + 5\kappa_{sat}),
\end{align}
while the isovector parameters are:
\begin{align}
 \nu^{IV}_{\alpha=0} &= E_{sym} - \frac59 t^{FG}_{sat} [1 + (\kappa_{sat}+3\kappa_{sym})],\\
 \nu^{IV}_{\alpha=1} &= L_{sym} - \frac59 t^{FG}_{sat} [2 + 5(\kappa_{sat}+3\kappa_{sym})],\\
 \nu^{IV}_{\alpha=2} &= K_{sym} - \frac{10}{9} t^{FG}_{sat} [-1 + 5(\kappa_{sat}+3\kappa_{sym})],\\
 \nu^{IV}_{\alpha=3} &= Q_{sym} - \frac{10}{9} t^{FG}_{sat} [4  + 5(\kappa_{sat}+3\kappa_{sym})],\\
 \nu^{IV}_{\alpha=4} &= Z_{sym} - \frac{40}{9} t^{FG}_{sat} [-7 + 5(\kappa_{sat}+3\kappa_{sym})].
\end{align}

In the previous implementations of MM the high density EOS was explored
through the application of MM to NSs \cite{Margueron2}, and the region 
around saturation when describing finite nuclei \cite{Chatterjee2017}.
In this study, we will concentrate on the sub-saturation density region
where we expect the crust-core (CC) phase transition in NS outer layer 
to occur. For this new application, we have to make sure that the MM method 
is still sufficiently flexible to successfully reproduce different existing 
models not only concerning the EoS, but also the spinodals, which are given 
as the second order derivatives of the energy density. This point will be 
discussed in Section \ref{subsec:Bparam}.

\subsection{Thermodynamical and dynamical spinodal}
 The CC phase transition can be seen as a transition from homogeneous matter 
 (the core) to clusterized one (the crust). Therefore, for a fixed temperature, 
 a region of instability occurs in which homogeneous matter is unstable towards 
 density fluctuations of finite linear size, fluctuations that are spontanously
 amplified to trigger the phase transition. This so-called spinodal region is 
 determined by the fluctuation free-energy curvature, i.e. by the unstable points 
 for which the variation of the free-energy surface following a fluctuation is 
 convex in at least one direction in the density space. Variation of free energy 
 can be written in a matrix form in three-dimensional space of density fluctuations 
 $\tilde{\Delta} = (\delta n_n, \delta n_p, \delta n_e)$ as 
 $\delta f = \tilde{\Delta}^{\ast} \mathcal{C}^f \tilde{\Delta}$, where $f$ is the 
 free energy density and $C^f$ is the free energy curvature matrix (see \cite{Ducoin2007} 
 for details). With the use of $\mu_i = (\partial f / \partial n_j)|_{T, n_{i \neq j}}$, 
 where $\mu$ is chemical potential, the curvature matrix is given by
 \begin{widetext}
 \begin{equation}\label{eq:Cmatrix}
 \mathcal{C}^f =\begin{pmatrix}
                \partial \mu_n / \partial \rho_n & \partial \mu_n / \partial \rho_p & 0\\
                \partial \mu_p / \partial \rho_n & \partial \mu_p / \partial \rho_p & 0\\
                0 & 0 & \partial \mu_e / \partial \rho_e
               \end{pmatrix}
               + 2 k^2 \begin{pmatrix}
                      C_{fin} + D_{fin} &  C_{fin} - D_{fin} & 0\\
                      C_{fin} - D_{fin} &  C_{fin} + D_{fin} & 0\\
                      %C^f_{nn} &  C^f_{np} & 0\\
                      %C^f_{pn} &  C^f_{pp} & 0\\
                      0 & 0 & 0
                     \end{pmatrix}
                     + \frac{4 \pi^2e^2}{k^2}\begin{pmatrix}
                                              0 & 0 & 0\\
                                              0 & 1 & -1\\
                                              0 & -1 & 1
                                             \end{pmatrix}.
 \end{equation}
 \end{widetext}

 The first term of eq. (\ref{eq:Cmatrix}) represents the free energy curvature 
 of a homogeneous system of neutrons, protons and electrons with respect to macroscopic 
 (infinite size) fluctuations. Matter must be neutral at a macroscopic scale to avoid a 
 divergence in the Coulomb energy, which sets the electron density equal to the proton 
 density, $n_e = n_p$.  This term alone determines the instability point of neutral 
 nuclear matter with respect to the nuclear liquid-gas (LG) phase transition, that is,
 the thermodynamical spinodal of the system. 

 It is however known that the LG point is only a qualitative estimation and that a better 
 description can be achieved by additionally considering the instability of stellar matter
 against the finite-size density fluctuations, that is, against the clusterization of matter. 
 This effect is taken into account by the additional, $k$-dependent terms in eq. (\ref{eq:Cmatrix}), 
 where a decomposition in plane waves is done, and $k$ is the wave number representing the size 
 of the density fluctuations, $k =  2\pi / \lambda$. 
 
 In the second term of eq. (\ref{eq:Cmatrix}), proportional to $k^2$, the coefficients 
 $C_{fin}$ and $D_{fin}$ are the isoscalar and isovector surface parameters, respectively.
 These parameters naturally appear as the lowest order parameters in a gradient expansion of
 the energy density functional around the expression for homogeneous matter 
 eq.(\ref{eq:en_particle}):
\begin{align}
\epsilon[n_0,n_1] =&
 e(n_0,n_1)  n_0 \nonumber \\ 
&+  C_{fin} 
 \left( \boldsymbol{\nabla}n_0 \right)^2
+  D_{fin} 
 \left( \boldsymbol{\nabla}n_1 \right)^2.
\label{eq_sym_density_energy_Skyrme}
\end{align}

For  Skyrme interactions, the surface parameters are defined straightforwardly in
terms of the non-local Skyrme force parameters $t_1, t_2, x_1$ and $x_2$
%through $C_{ij}$ parameters  ($i,j = n, p$) in the following manner
% \begin{eqnarray}\label{eq:C_Skyrme}
% C_{fin} + D_{fin} &=& C_{nn} = C_{pp} \\
% C_{fin} - D_{fin} &=& C_{np} = C_{pn}.
% \end{eqnarray}
% The relations between the two representations defined in eq. \ref{eq:C_Skyrme} are obtained through a simple equalization of surface instability Hamiltonian 
% \begin{align}
% C_{nn}|\nabla \rho_n|^2 +& C_{pp}|\nabla \rho_p|^2 + 2C_{np}|\nabla \rho_n\ \nabla \rho_p| \nonumber \\
% &=C_{fin}|\nabla \rho|^2 + D_{fin}|\nabla (\rho_n - \rho_p) |^2 .
%\end{align}
% For Skyrme interactions their values are defined  with the Skyrme force parameters
 \begin{eqnarray}
 C_{fin} + D_{fin} &=& \frac{3}{16} [ t_1 (1-x_1) - t_2 (1 + x_2) ], \\
 C_{fin} - D_{fin} &=& \frac{1}{16} [ 3 t_1 (2 + x_1) - t_2 (2 + x_2) ]
\end{eqnarray}
 as given in, for example, Ref. \cite{CHABANAT1998231}. 
 % Debi asks to define t1, t2 etc? I think reference is enough?

In case of meta-modeling, $C_{fin}$ and $D_{fin}$ are added as two extra parameters to our 
empirical EoS parameter set. They are constrained from experimental nuclear masses, as 
will be explained in Section \ref{sec:Sensitivity}.

 %or in general for any EOS whose empirical parameters are known, the values of $C_{fin}$ 
 %and $D_{fin}$ are obtained through the $\chi^2$ minimization  procedure as it will be 
 %described in section \ref{sec:Sensitivity}. 

 The  final term in eq. (\ref{eq:Cmatrix}), proportional to $1/k^2$, comes
 from the Coulomb interaction induced by the plane-wave charge distribution. In the 
 limit $k \rightarrow 0$ the thermodynamic fluctuations are recovered.

% For finite nuclei where $Z \sim N$, the isovector term coming from the difference 
%between the neutron and proton densities can be neglected and,  with setting $D_{fin} = 0$, 
%the four coefficients in second term of matrix  (\ref{eq:Cmatrix}) are the same and represented 
%with $C_{fin}$ only as it was done in finite nuclei study through metamodeling method \cite{Chatterjee2017}.
%In that case, the $C_{fin}$ value was found by the $\chi^2$ fit of binding energies for several symmetric
%nuclei, being equal to $C_{fin} = 59 \pm 13$ $MeV fm^{-5}$.  In the present study though, when we explore 
%the CC transition, the matter is very asymmetric and the isovector surface parameter $D_{fin} \neq 0$.
%This means that the  inclusion of finite-size density fluctuations will bring in two additional parameters, 
%$C_{fin}$ and $D_{f}$. 
 
 Stable matter correponds to a positive curvature in each point  of the $n_n - n_p$ space. 
 This is equivalent to requiring that the eigenvalues of the curvature matrix, $\lambda_{\pm}$
 are positive. It turns out that  $\lambda_{+}$ is always positive, reflecting the fact that
 the order parameter of the transition is a scalar, while $\lambda_{-} \rightarrow 0$ at the spinodal 
 boundary and becomes negative inside the instability region. Finding the spinodal curve is thus 
 equivalent to finding the $\lambda_{-} = 0$ solutions for the full $n_n - n_p$ space. In case of 
 dynamical spinodal, as it is shown in Ref. \cite{Ducoin2007}, there is a range of $k$ values with
 $\lambda_{-} < 0$ with lower boundary having $k \sim 20$ $MeV$ and the upper boundary being
 $k \sim 200$ $MeV$ (depending on the EoS model). By the additional variation of the wave number
 $k$ between these values for each point of the $n_n - n_p$ space the $k$-envelope is found, which 
 defines the dynamical spinodal, that is the density region where matter is unstable with respect 
 to density fluctuations of at least one given $k$. 
 
 The crossing between the dynamical (thermodynamical) spinodal and the
 $\beta$-equilibrated EOS defines the CC (LG) phase transition point that 
 we are looking for. 
 
\section{The spinodal calculation within the MM approach}\label{sec:MMspinodal}

\subsection{The role of $\bf{N>2}$ order parameters in the sub-saturation density region}

As we go further from the saturation density, the higher orders in the Taylor 
expansion of the energy per particle around the saturation density become
more relevant. For the finite nuclei study of Ref. \cite{Chatterjee2017}, 
the expansion up to $N=2$ was sufficient for a good reproduction of experimental
observables, while for the study of NSs \cite{Margueron2} the expansion up 
to order $N=4$ was shown to be necessary.

\begin{figure}[t!]
    \centering
    \includegraphics[scale=0.33]{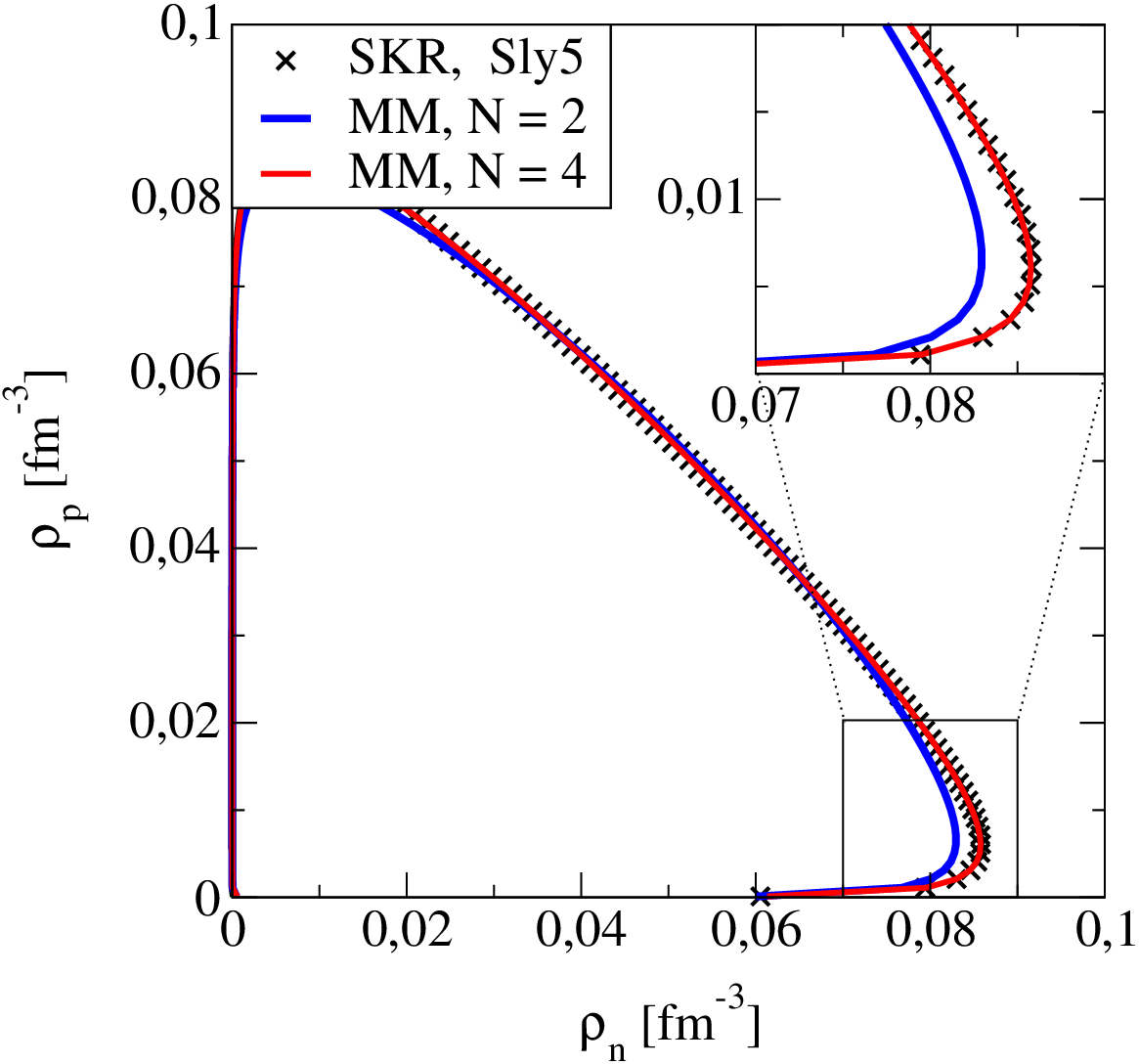}
    \caption{(Main plot) The thermodynamical spinodal for the Skyrme Sly5 model. 
    SKR stands for the original Sly5 model spinodal while MM marks the spinodals calculated
    in the MM framework, for $N=2$ and $N=4$ cases. (Inserted plot) Enlarged neutron rich part
    of the spinodal where the CC transition point is to be found.}
    \label{fig:N2N4}
\end{figure}

The effect of taking higher orders into account in the sub-saturation density 
region can be explored through the study of spinodals, the instability region
at low density that defines the crust-core phase transition point in the NS
outer layer. As an example, the spinodal calculation of Sly5 Skyrme interaction 
model is given in Figure \ref{fig:N2N4}. The shape of original thermodynamical 
Sly5 spinodal, calculated via the Skyrme functional and marked with SKR, is 
compared to the MM calculations using the same empirical parameters as in Sly5, 
for the $N=2$ and $N=4$ cases. From  Figure \ref{fig:N2N4} it is clear that the 
higher order parameters (up to $N=4$) play a non negligible role in this 
density region and will therefore be taken into account.

\subsection{Low density correction parameter b} \label{subsec:Bparam}

In  previous studies using the meta-modeling method \cite{Margueron1,Margueron2,Chatterjee2017} 
the low density correction was  defined fixing the $b$ parameter entering equation (\ref{eq:b_exp}) 
as %equal to $1/2$. This is achieved by the appropriate choice of $n_0$, the density at which the
%low density correction is fixed, that defines the value of $b$ parameter through
$b = \frac{n_{sat}}{n_m}ln(2)$,  a quick vanishing of the correction being secured by $n_m$ being
small but finite. We recall that the physical meaning of $n_m$ is the minimum density at which the 
Taylor expansion around saturation is considered to be valid. Since the correction term plays a role
only at low densities, a constant choice for $b$ or equivalently $n_m$ will not influence the result
when MM is applied to finite nuclei \cite{Chatterjee2017} or NSs \cite{Margueron2} as long 
as the value of $n_m$ is kept well below the saturation density. Nonetheless, the question is if this
is true when we explore the sub-saturation region, in search for the crust-core phase transition point 
expected in the density range $0.05$ $fm^{-3} < n_n < 0.09$ $fm^{-3}$ \cite{Ducoin:2011fy}, and with $n_p$ an order of magnitude lower. It is 
therefore worthwhile to test the effect that a variation of the $b$ parameter will have on the spinodal 
shape for very asymmetric nuclear matter, where we expect to find the CCPT point. Several different 
values of $b$ parameter are given in the Table \ref{tab:b_values}, as well as the corresponding density 
$n_m$ at which the low density correction is fixed.

\begin{table}[t]
  \centering
   \bgroup
\def\arraystretch{1.3}
  \caption{ Change of the b parameter with $n_m$, the density at which the
  low density correction is fixed (see text). \\}
  \label{tab:b_values}
  \begin{tabular}{l c c c c c r}
  \hline
  \hline
    &\multicolumn{1}{c}{$n_m$} & & $n_m [fm^{-3}]$ & & $b$ &\\ %= \frac{n_{sat}}{n_m}ln(2)$ \\%& meta-model \\
  \hline
    & $0.075$ $n_{sat}$ & & $\sim 0.012$	& & $\sim 9.24$	&\\%& MM0\\    
    & $0.1$ $n_{sat}$   & & $\sim 0.016$	& & $\sim 6.93$	&\\%& M\M1\\
    & $0.2$ $n_{sat}$   & & $\sim 0.032$	& & $\sim 3.46$	&\\%& MM2\\
    & $0.3$ $n_{sat}$   & & $\sim 0.048$	& & $\sim 2.31$ &\\% MM3\\
    & $0.4$ $n_{sat}$   & & $\sim 0.064$	& & $\sim 1.73$ &\\%& MM4\\
    & $0.5$ $n_{sat}$   & & $\sim 0.08$		& & $\sim 1.38$ &\\%& MM5\\
    & $0.6$ $n_{sat}$   & & $\sim 0.096$  	& & $\sim 1.15$ &\\%& MM6\\
   \hline
   \hline
  \end{tabular}
  \egroup
 \end{table}

\begin{figure}[b!]
\centering
%%% == trim={<left> <lower> <right> <upper>}
%\includegraphics[scale=0.25, angle=0]{./figures/Sly5_MM.eps}
%\includegraphics [scale=0.15]{./Figure2a.eps}
\includegraphics[trim=0cm 0cm 0cm 0cm,clip=true,scale=0.18]{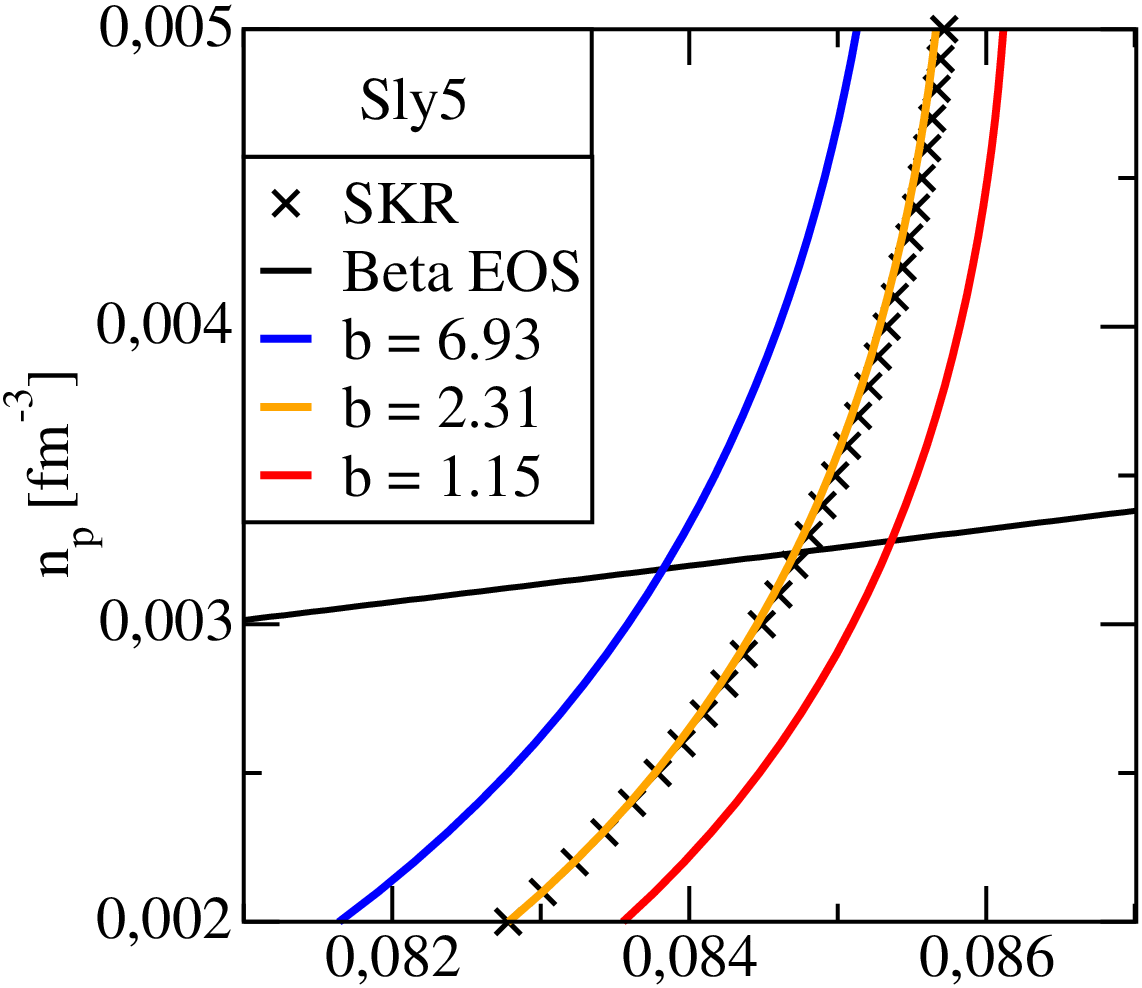} 
\includegraphics[trim=0cm 0cm 0cm 0cm,clip=true,scale=0.18]{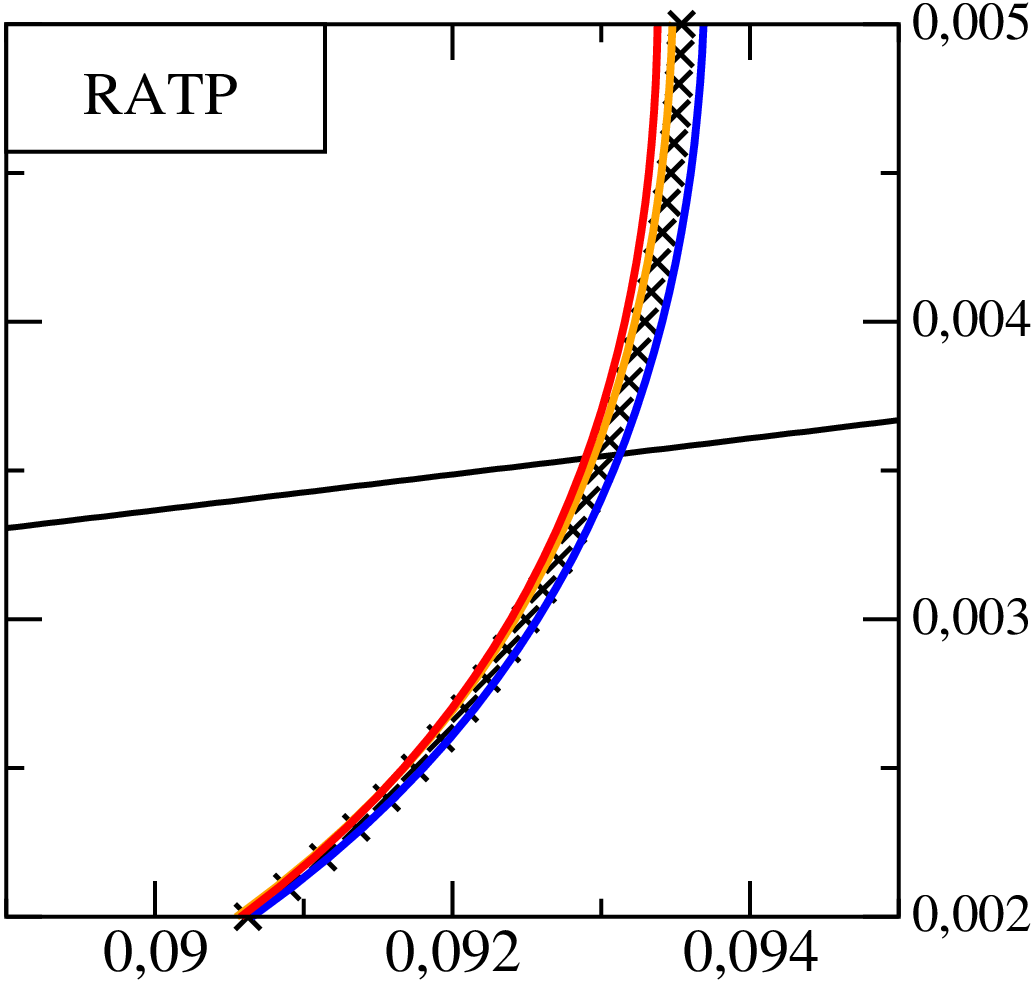} 
\includegraphics[trim=0cm 0cm 0cm 0cm,clip=true,scale=0.18]{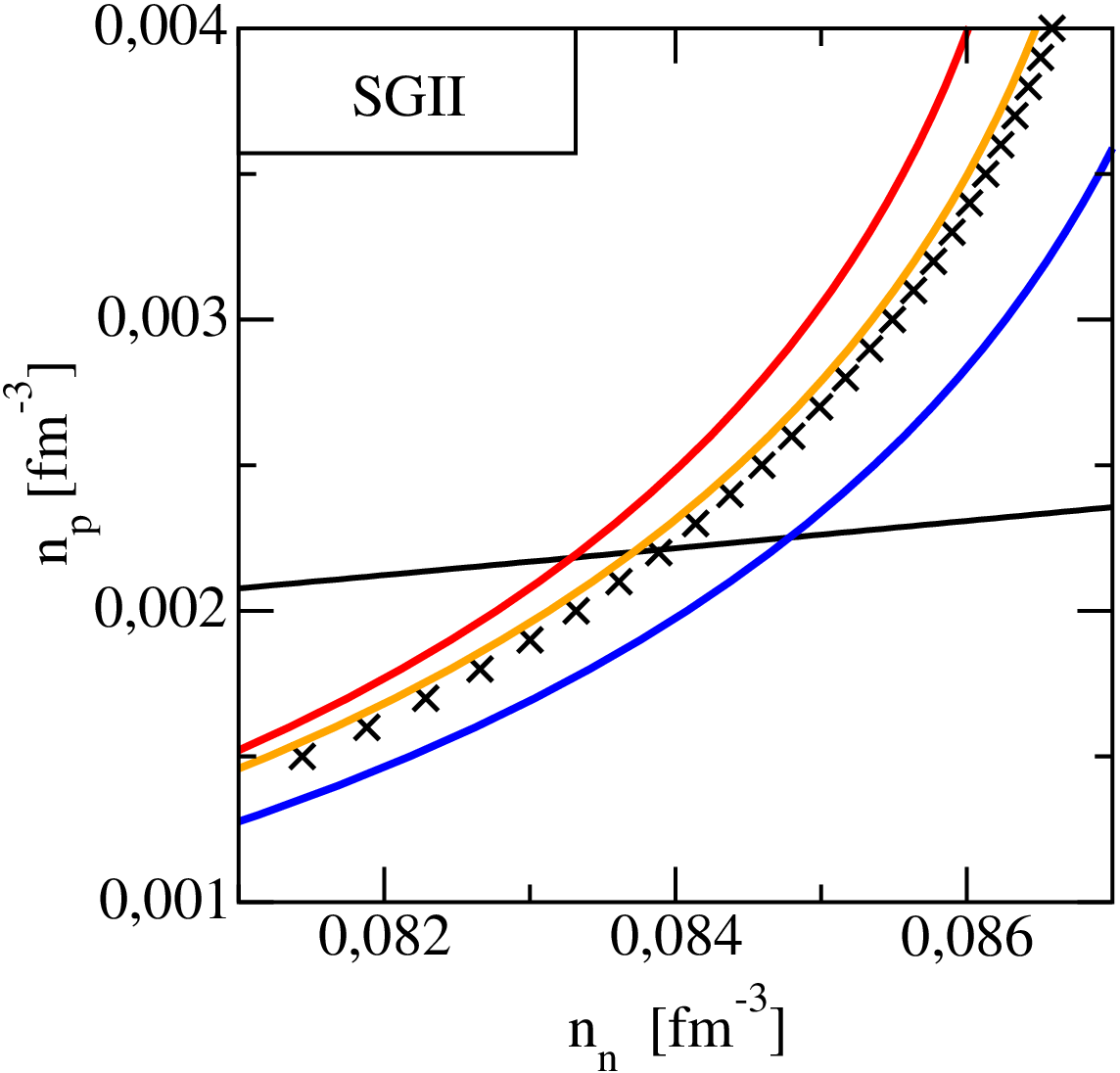} 
\includegraphics[trim=0cm 0cm 0cm 0cm,clip=true,scale=0.18]{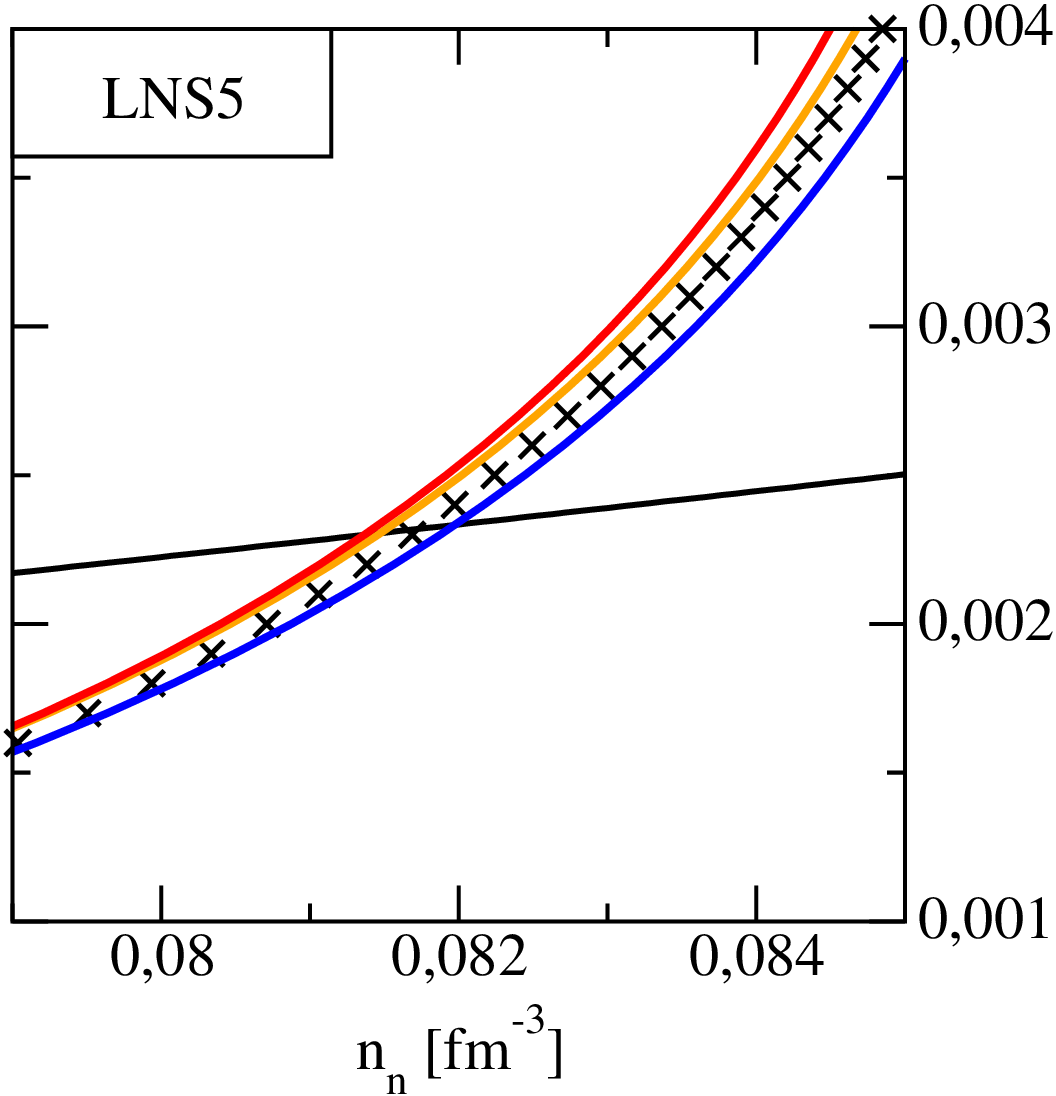} 
\caption{The influence of the low density correction parameter $b$ on the spinodal
shape (for $N=4$) within the MM method for four different models
(Sly5 \cite{CHABANAT1998231}, RATP \cite{RATP}, SGII \cite{vanGiai:1981zz}
and LNS5 \cite{LNS5}), compared to the original spinodal 
marked with SKR. See text for further details. }
\label{fig:b_influence}
\end{figure}

The densities above $n_m> 0.6$ $n_{sat}$ are not worthwhile to look at since we 
 want to fix the low density correction well below saturation. The physically reasonable $n_m$ span the
 $b$ value over the range $1 < b < 10$. We can explore the effect that the change of the
 b parameter value has on the spinodal shape, while keeping in mind that the low density
 correction should be fixed at the density below the expected CC transition point density. 
 This essentially means that the values of $b\leq2$ can be excluded, since these values
 correspond to the densities $n > 0.05$ $fm^{-3}$. The calculation of thermodynamical 
 spinodal shapes is done within the MM approach for four different models
 (Sly5 \cite{CHABANAT1998231}, RATP \cite{RATP}, SGII \cite{vanGiai:1981zz} %- whose parameters are optimized to describe nuclear collective motion, 
 and LNS5 \cite{LNS5}) and the result are compared with the exact  spinodal, as 
 given in Fig.\ref{fig:b_influence}. The exact spinodal of the different models
 is marked with SKR while the spinodals calculated in the MM approach are marked with different 
 $b$ values. The black line represents the stellar EOS whose crossing with spinodal gives the CC 
 transition point density $n_t = n_{n,t} + n_{p,t}$.
 
 The best reproduction of the original  models seems to be obtained for the 
 $b$ value $b \sim 2.31$ in all four cases. While the variation of
 $b$ changes the spinodal shape by different rate for the four chosen 
 interactions, %causing, 
%  for example, a very slight shift in $n_t$ for RATP model, while for Sly5 model
%  the difference is enhanced. In both cases however, 
 the maximal change is of the order of $n_t \pm 0.001$ $fm^{-3}$ for all studied
 cases, which is one order of magnitude below the expected uncertainty on the CC 
 point. Therefore, we can conclude that the variation of the  $b$ parameter will certainly 
 not give the dominant effect. It is important to point out that this is the consequence 
 of including the higher order parameters ($N=3,4$) in the meta-model framework.
 When the expansion parameters in eq. (\ref{eq:expansion}) are taken only up to $N=2$, 
 the change of the $b$ parameter in the low density region would cause more 
 significant change of the CC density, of order $n_t \pm 0.01$ $fm^{-3}$. This is a
 first indication that higher order parameters, beyond the usually studied slope of the 
 symmetry energy $L_{sym}$, must be considered.
 
\section{Sensitivity study of thermodynamical and dynamical spinodals}\label{sec:Sensitivity}

The parameters of the MM approach are summarized as:
\begin{itemize}
 \item empirical EOS parameters $\{P_{\alpha}\} \equiv \{n_{sat}$, $E_{sat}$, $K_{sat}$, $Q_{sat}$
 $Z_{sat}$, $E_{sym}$, $L_{sym}$, $K_{sym}$, $Q_{sym}$, $Z_{sym}\}$;
 \item $\kappa_{sat}$ and $\kappa_{sym}$ that parametrize the isoscalar Landau effective mass $m^\ast$ 
 as well as the proton-neutron mass splitting $\Delta m^\ast$;
 \item low density correction parameter $b$;
 \item for the case of dynamical spinodal, the isoscalar and isovector surface parameters
 $C_{fin}$ and $D_{fin}$ .
\end{itemize}

The question is how sensitive are the density and pressure of the phase transition point  
to the change of each of these parameters. To get an insight on this issue, two different
reference parameter sets were used for the parameter sensitivity study of the LG and CC point,
listed in table \ref{tab:Tab46Tab7}. The empirical EOS parameters $\{P_{\alpha}\}$ and the 
parameters $m^{\ast}$ and $\Delta m^{\ast}$ belonging to Set 1 in table \ref{tab:Tab46Tab7} 
are obtained by averaging the single parameter value and its associated uncertainties over a
total of 50 different Skyrme, RMF and RHF models, which all were successfully compared to 
experimental nuclear data. The low order (LO) parameters up to second order ($N=2$) are
taken from table IV of reference \cite{Margueron1}. The higher order (HO) parameters 
($N=3,4$) are taken from table VI of the same reference where HO parameters are
better constrained in the MM approach by fixing the EOS at an additional reference
point in density, $n_0 = 4 n_{sat}$, besides the saturation density. 

A second choice for a reference parameter set (Set 2 in table \ref{tab:Tab46Tab7})
is taken from table VII of \cite{Margueron1} and it contains LO parameters (up to $N=2$) 
extracted from experimental analysis, while the $Q_{sat,sym}$ and $Z_{sat,sym}$ are 
estimated from table VI, as for Set 1. 

\begin{table*}[t!]
 \centering
 \bgroup
\def\arraystretch{1.3}
 \caption{The sets of empirical parameters and its associated uncertainties for the two EOS model 
 reference points taken from reference \cite{Margueron1}. The additional parameters $b$, $C_{fin}$ 
 and $D_{fin}$ are chosen or calculated (see text for details).}
 \label{tab:Tab46Tab7}
 \begin{tabular}{ccccccccccccccccc}
 \hline
 \hline
    Parameter & & $n_{sat}$ & $E_{sat}$ & $K_{sat}$ & $Q_{sat}$ & $Z_{sat}$ & $E_{sym}$ & $L_{sym}$ 
    & $K_{sym}$ & $Q_{sym}$ & $Z_{sym}$ & $\frac{m^{\ast}}{m}$ & $\frac{\Delta m^{\ast}}{m}$ & $b$ & $C_{fin}$ & $D_{fin}$ \\
       set & & $[\frac{1}{fm^{3}}]$  & $[MeV]$  & $[MeV]$  & $[MeV]$  & $[MeV]$ & $[MeV]$ & $[MeV]$ 
    & $[MeV]$ & $[MeV]$ & $[MeV]$ &  &  & $[MeV]$ & $[\frac{MeV}{fm^{5}}]$ & $[\frac{MeV}{fm^{5}}]$ \\
\hline
    Set 1  & & $0.1543$  & $-16.03$ & $251$ & $287$ & $-1765$ &
 	     $33.30$   & $76.6$   & $-3$  & $-124$& $-437$  & $0.72$ & $0.01$ & $2.31$ & $52$ & $-30$\\
 	     
$\sigma_1$ & & $ 0.0054$ & $ 0.20$ & $ 29$ & $ 352$ & $ 782$ &
	     $ 2.65$   & $ 29.2$ & $ 132$& $ 317$ & $ 594$ & $ 0.09$& $ 0.2$ & $2$ and $10$& $ 52 $ & $ 30$ \\
 
 Set 2   & & $0.155$  & $-15.8$ & $230$  & $300$ & $-500$ &
             $32$     & $60$    & $-100$ &$0$    & $-500$ & $0.75$ & $0.1$ & $2.31$ & $46$ & $-70$ \\
           
$\sigma_2$ & &$ 0.005$& $ 0.3$ & $20$ & $400$ & $1000$ &  
           $ 2$ & $15$ & $ 100$ & $400$ &  $1000$ & $0.1$ & $ 0.1$ & $2$ and $10$ & $46$& $ 70$\\
 \hline
 \hline
 \end{tabular} 
 \egroup
\end{table*} 
The average value for the $b$ parameter is taken to be $b=2.31$, the value that reproduces the four chosen  model spinodals best. To study its sensitivity, this parameter is varied between the 
values of $2<b<10$, which matches the fact that the low density correction is fixed at
a density between $ 0.01 fm^{-3} < n_0 < 0.05 fm^{-3}$.

The additional $C_{fin}$ and $D_{fin}$ parameters governing the surface properties of the energy
functional are optimized for each EoS parameter set on experimental nuclear masses taken from the 
AME2012 mass table \cite{AME2012}. 

Given a set of empirical parameters, the binding energies of several symmetric spherical nuclei
($^{40}Ca$, $^{48}Ca$, $^{48}Ni$, $^{58}Ni$, $^{88}Sr$, $^{90}Zr$, $^{114}Sn$, $^{132}Sn$, $^{208}Pb$) 
are calculated in the extended Thomas Fermi (ETF) approximation using the parametrized density profiles 
of Ref. \cite{Aymard2014}. $C_{fin}$ and $D_{fin}$ are obtained from the following chi-square minimization:
\begin{equation}
 \chi^2 = \sum^N_{i=1} (E_{i,th} - E_{i, exp})^2 / E_{i,exp}^2.
\end{equation} 
The optimal values, $C^{(opt)}_{fin}$ and $D^{(opt)}_{fin}$, for Set 1 and Set 2 that give
minimal $\chi^2$ are given in table \ref{tab:Tab46Tab7}, with the $C_{fin}$ value being
close to the previously found result for a study of finite nuclei \cite{Chatterjee2017}. 

To study the sensitivity of the CC point to the finite size parameters, a physical interval
has to be specified reflecting the uncertainty around the optimized value. An estimation for
this interval can be obtained if we consider that the dynamical spinodal (corresponding to 
finite values for $C_{fin}$,$D_{fin}$) is always enveloped by the thermodynamical one 
(corresponding to $C_{fin}$=$D_{fin}=0$) . Imposing this condition gives the approximate 
ranges of the parameters $0 \lesssim C_{fin} \lesssim100$ and $0 \lesssim D_{fin} \lesssim -150$, 
as given in table \ref{tab:Tab46Tab7}.%, for which the case of two parameters being equal to zero represents
%the thermodynamical spinodal case.

\begin{figure}[t] 
\centering
\includegraphics[scale=0.33, angle=0]{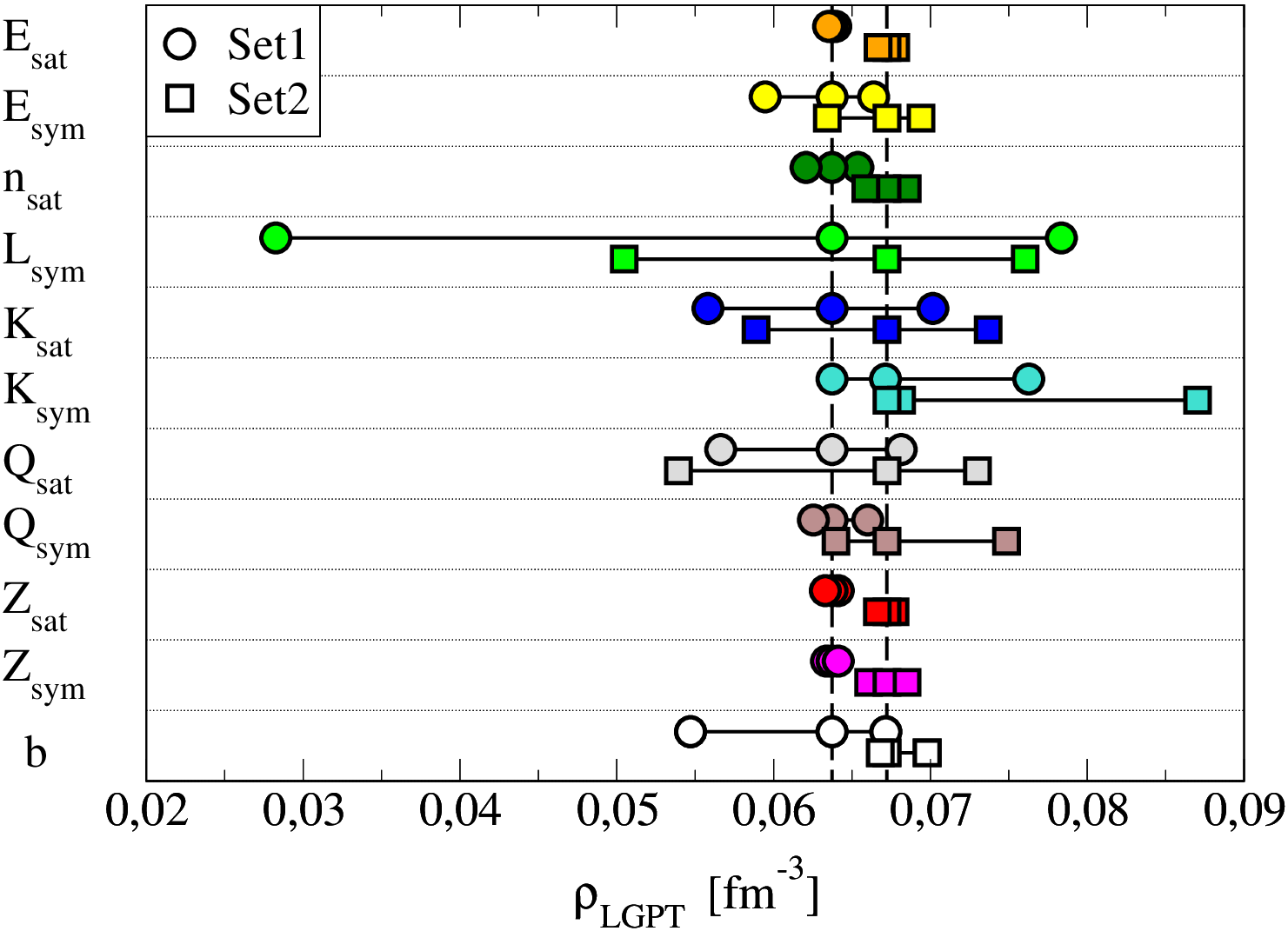}\\
\vspace{5mm}
\includegraphics[scale=0.33, angle=0]{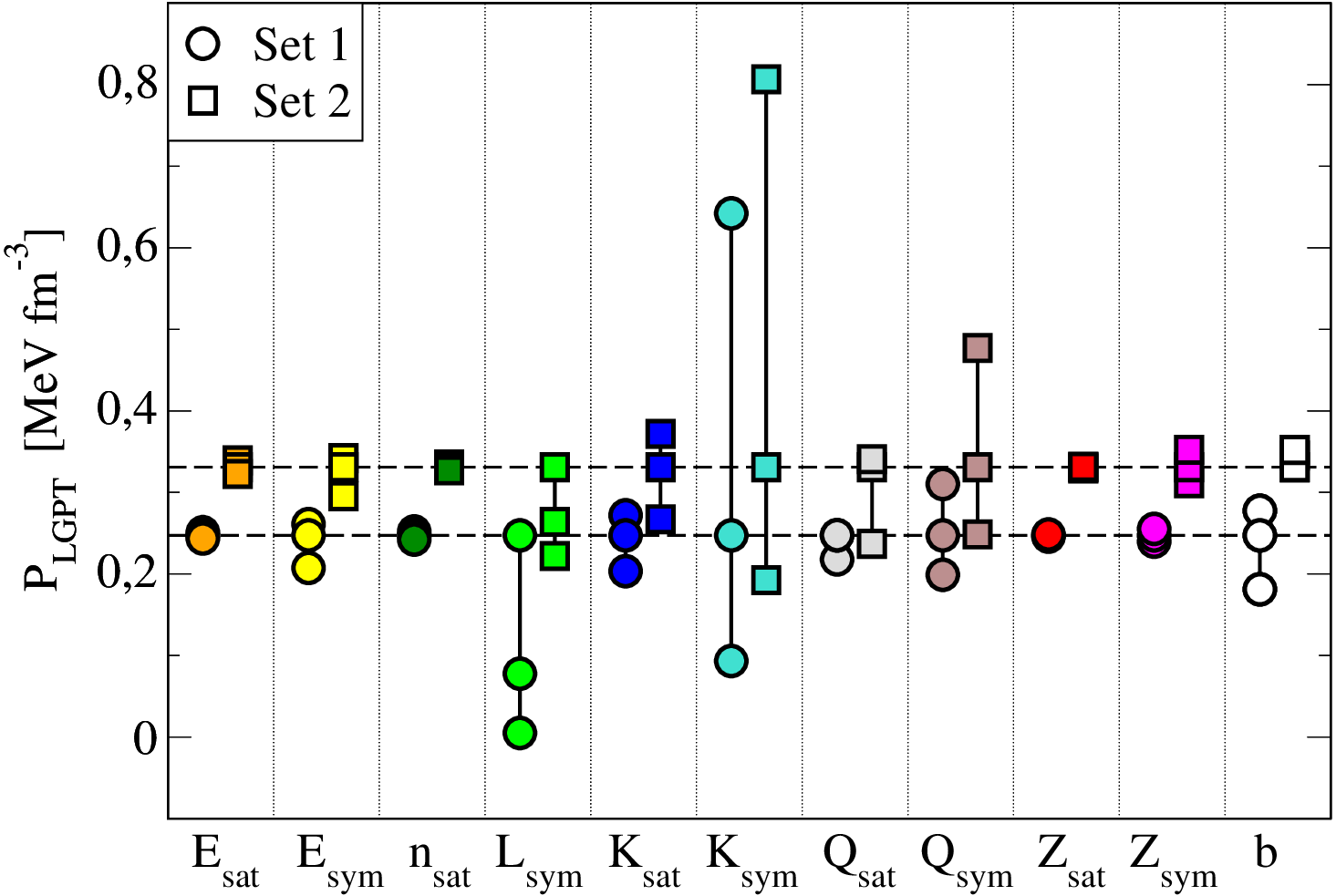}
\caption{Sensitivity parameter study of the LG phase transition (LGPT) point for two 
reference points with parameters sets Set 1 and Set 2 (table (\ref{tab:Tab46Tab7}))
taken from Ref. \cite{Margueron1}. The plots represent the sensitivity of
transition density $n_t$ (Top) and pressure $P_t$ (Bottom) on the change of 
parameter values for $av \pm \sigma$.}
\label{fig:LGPT}
\end{figure}

\begin{figure}[t]
\centering
\includegraphics[scale=0.33, angle=0]{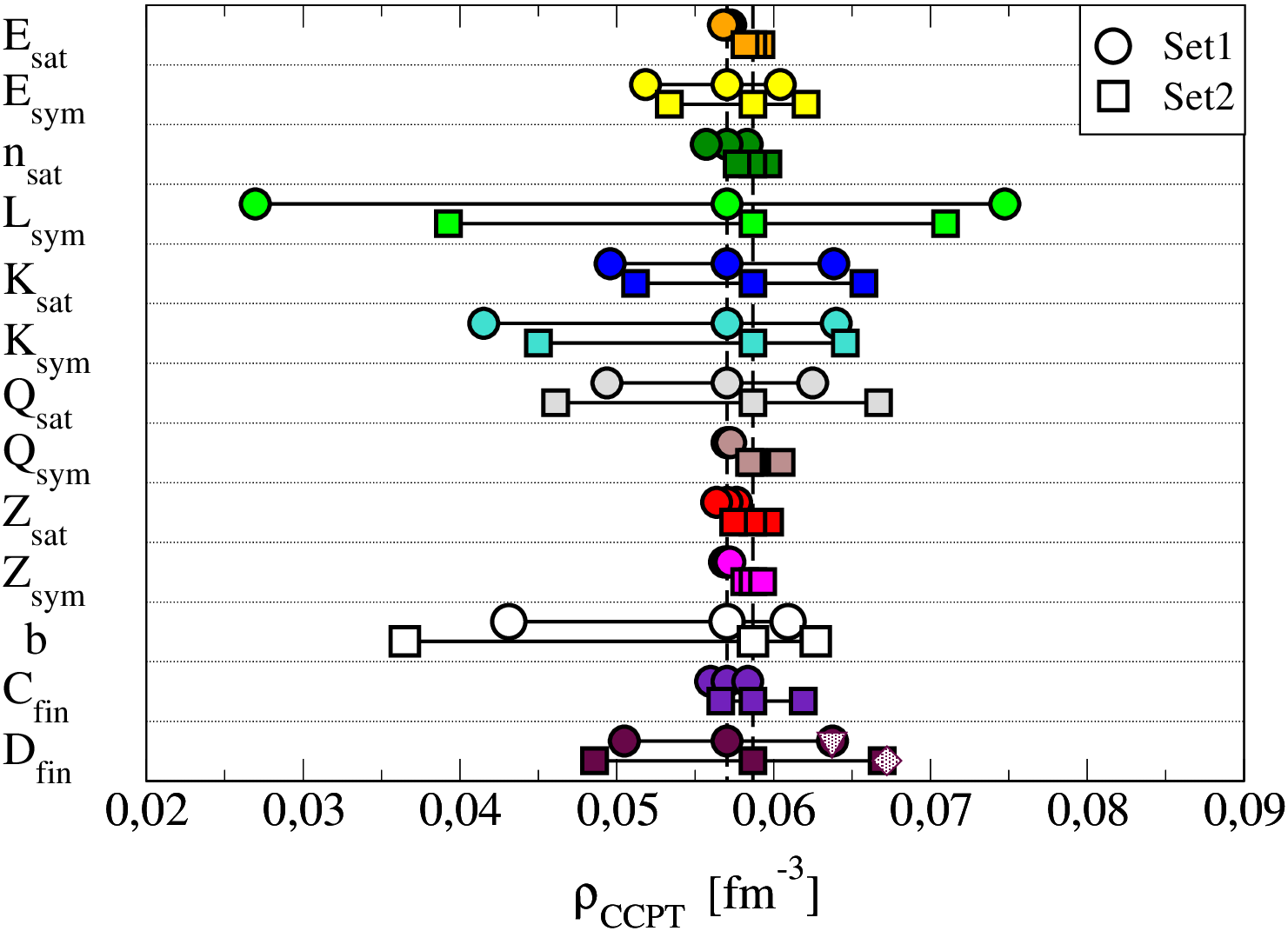}\\
\vspace{5mm}
\includegraphics[scale=0.33, angle=0]{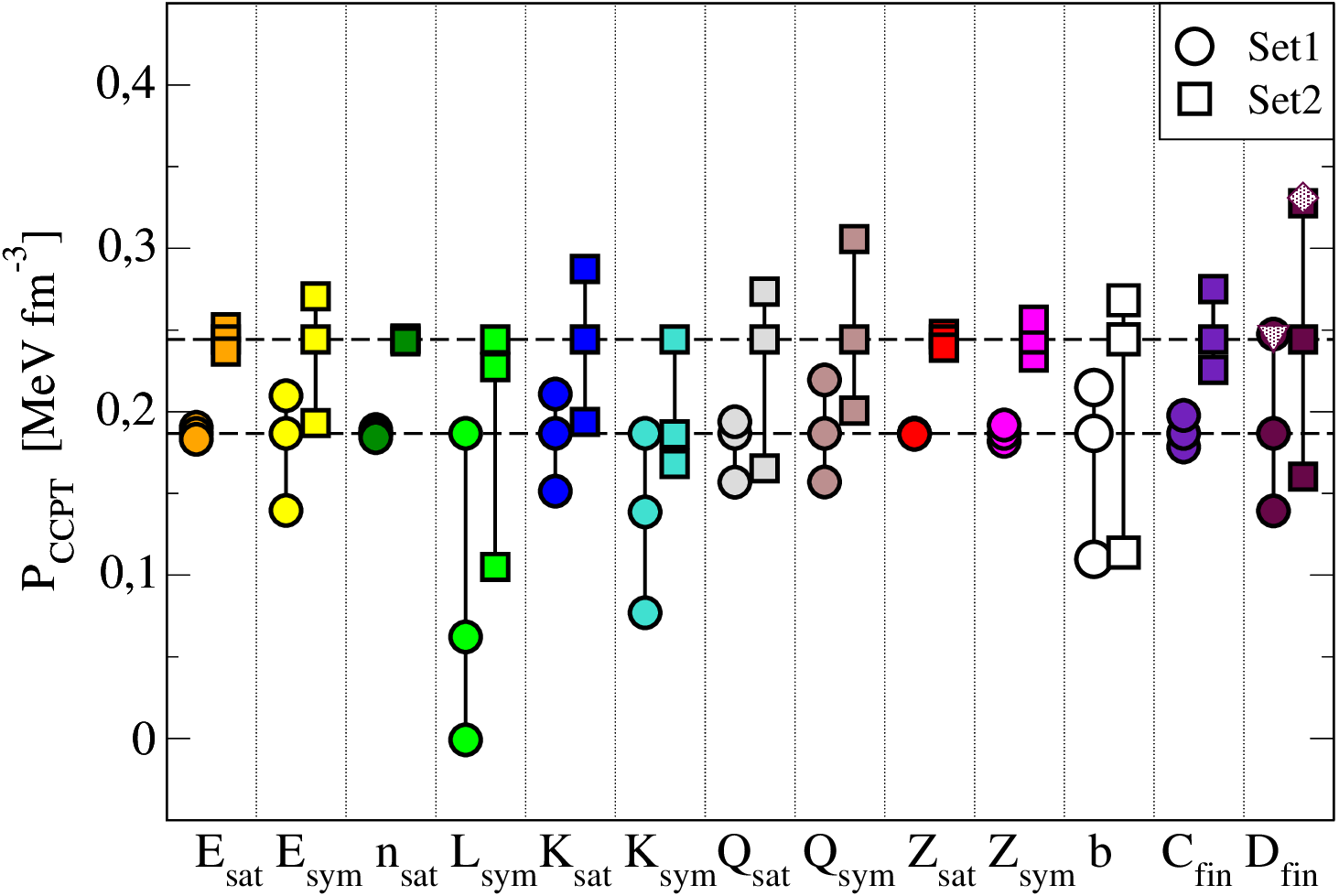}
\caption{Same as Fig. \ref{fig:LGPT} but for the CC phase transition (CCPT) point. 
Two additional parameters that come into play, $C_{fin}$ and
$D_{fin}$, describe the surface properties of the functionals.\vspace{7mm}}
\label{fig:CCPT}
\end{figure}

At this point, once we have the full sets of parameters ($\{P_\alpha\},m^\ast,\Delta m^\ast, b, C_{fin}, D_{fin}$)  
according to set 1 and set 2, we find the thermodynamical (dynamical) spinodal of the system 
and, by finding the crossing between the spinodal and the $\beta$-equilibrated EOS, obtain the
LG (CC) transition point density $n_t$ and pressure $P_t$. Further, by changing each of the 
parameter values individually by associated uncertainty $\pm \sigma$ we get an idea of the 
transition point sensitivity to each of the parameters individually. This sensitivity study is
enveloped in Figure \ref{fig:LGPT} for the thermodynamical spinodal and in Figure \ref{fig:CCPT} 
for the dynamical one. For both figures, the top panels give the sensitivity of the density $n_t$ 
while the bottom ones represent the sensitivity of pressure $P_t$. The two dashed lines represent
the transition value (in density or pressure) for the given sets from table \ref{tab:Tab46Tab7}, 
Set 1 being marked by circles and Set 2 by squares. Set 2, representing the parameter values
constrained by experiment, gives slightly higher values for both transition density and pressure.
There are three points depicted for each parameter of the two sets: the average value, which sits
on the dashed line, and the two average $\pm \sigma$ values in addition.

In both cases, from the range of values for each of the parameters, we observe that 
the parameters $L_{sym}$ and $K_{sym}$ have the highest influence on LG and CC points 
both in transition density and pressure. If the influence of $L_{sym}$ was already reported
in previous studies \cite{Ducoin:2011fy}, the effect of $K_{sym}$ is less known.
The reader should keep in mind that the very wide bands, as e.g. for $L_{sym}$ parameter
for Set 1, are given by the average and uncertainty of $L_{sym}$ over 50 different models
which therefore results in very wide range of values, $47.4 < L_{sym} < 105.8$. In contrast
to that, the values from Set 2, extracted from dedicated experimental studies, are more 
restrictive, being $L_{sym} = 60 \pm 15$. The same goes for all the parameters.

% Let's point out that the range of $n_t$ and $P_t$ obtained for $b$ parameter is taken over 
% the explored region from figure \ref{fig:b_influence}. Even thought it seems that 
% it has a high influence on transition point, especially for the dynamical spinodal case,
% there is a good reason to restrict this parameter more firmly, if not even to fix its value 
% as discussed in subsection \ref{subsec:Bparam}, while we can not say the same for the 
% $L_{sym}$ and $K_{sym}$ parameters.
%{\bf I do not understand the above paragraph. I would rather say: 
It is also interesting to observe the very high sensitivity of the CC point to the low density
parameter $b$. Even if the range for this parameter has been fixed in a largely arbitrary way 
and the functional form for the low density correction is also largely arbitrary, the observed
sensitivity reflects the fact that the transition point depends on the properties of the 
functional in the very low density region, where the Taylor expansion breaks down because 
higher order correlation beyond the mean field might affect the the functional form beyond 
the derivatives at the saturation point.

\begin{table}[b!]
  \centering
   \bgroup
\def\arraystretch{1.3}
  \caption{The phase transition density and pressure  of different reference models 
  (Set 1 and Set 2) for the LGPT and CCPT cases.\\}
  \label{tab:Pn_Values}
  \begin{tabular}{c c c c c c c c c}
  \hline
  \hline
    &  & \multicolumn{3}{c}{ $n_t$ }& & \multicolumn{3}{c}{$P_t$} \\ 
    &  &LGPT & &CCPT & & LGPT& & CCPT \\%& meta-model \\
  \hline
    Set 1 & & $0.0637$& &$0.0570$ & & $0.2472$ & &$0.1867$   \\
    Set 2 & & $0.0672$& &$0.0587$ & & $0.3309$ & &$0.2442$   \\
   \hline
   \hline
  \end{tabular}
  \egroup
 \end{table}
The transition points obtained for the two cases are given in table \ref{tab:Pn_Values}.
As expected, when finite-size density fluctuations are included, the transition point 
is found at somewhat lower density and pressure.
 
\section{Bayesian analysis of the dynamical spinodal CCPT point}\label{sec:Bayesian}

In the next step, we restrict ourselves to the dynamical spinodal which represents more accurately
the real transition point, and fully explore the parameter space of the meta-modeling, 
up to fourth order in Taylor expansion ($N=4$), within a Bayesian analysis. The prior is 
defined as a flat distribution of each of the parameters
$(\vec X) =(\{P_\alpha\},m^\ast,\Delta m^\ast, b)$, with a parameter values interval 
given in table \ref{tab:param}.

The posterior distribution is obtained by applying two different physical filters to 
the prior distribution,
\begin{equation}
p_{post} (\vec X) = \mathpzc{N} \, w(\vec X) \, e^{-\chi^2(\vec X)/2} \, p_{prior}(\vec X)  .
\label{eq:probalikely}
\end{equation}
In this expression, both strict ($w$ term) and likelihood (exponential term) filters are
applied, and  $\mathpzc{N}$ is a normalization. $\chi^2(\vec X)$ represents the $\chi^2$
corresponding to the optimal fit of nuclear masses which is done to determine the surface
tension parameters $C_{fin}$ and $D_{fin}$ for each $\vec X$ parameter set (see section
\ref{sec:Sensitivity}). $w$ is a sharp $\delta$-function filter that outputs 1 if the 
constraint is respected, and 0 otherwise. The constraint consists in requiring that each 
model lays within the uncertainty band of the N3LO effective field theory calculation of
energy and pressure for SM and pure neutron matter (PNM) of Ref. \cite{Drischler2016}
in the density interval $0.05 fm^{-3} < n < 0.2 fm^{-3}$.

In this global bayesian determination of the uncertainty intervals for the a-priori independent 
model parameters, the determination procedure of $C_{fin}(\vec X)$ and $D_{fin}(\vec X)$ by $\chi^2$ 
minimization can be only justified if the $\chi^2$ minimum is sufficiently shallow to consider 
that any other value for $C_{fin}$, $D_{fin}$ would be filtered out once the likelihood constraint
on nuclear masses is applied. 

\begin{figure}[t!]
\centering
\includegraphics[scale=0.33, angle=0]{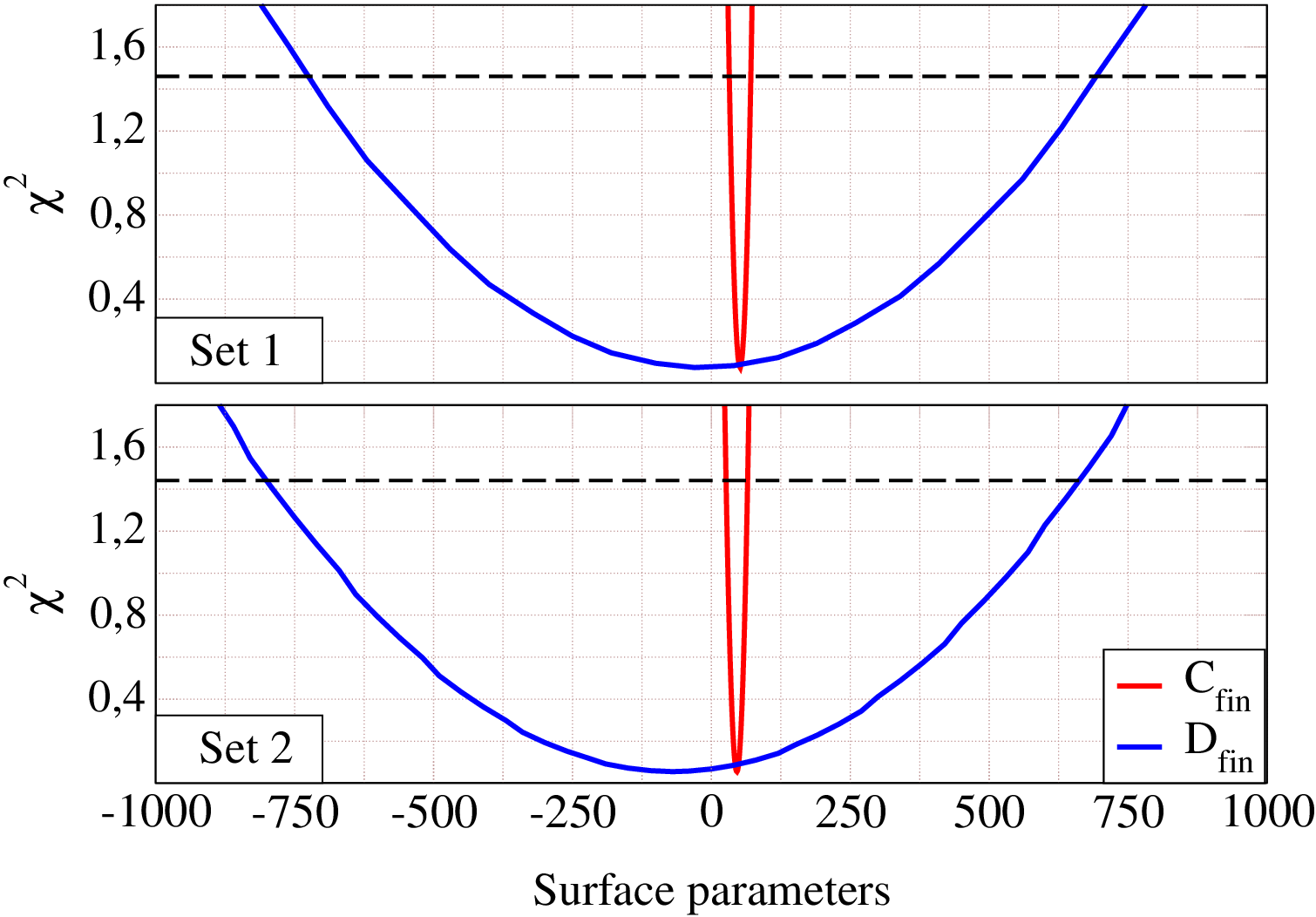}
\caption{The $\chi^2$ minimum for $C_{fin}$ and $D{fin}$ parameters for the two 
reference point models.}
\label{fig:CfinDfin}
\end{figure}

To explore this issue, we consider the distribution of $\chi^2$ value around minimum for two
representative reference models, shown in Figure \ref{fig:CfinDfin}. The behavior is qualitatively
similar for all meta-modeling parameter sets.

We can see that the $D_{fin}$ minimum is very flat, reflecting the fact that stable nuclei very 
poorly constrain the isovector gradient terms. For that reason, we enlarge our prior distribution
by associating to each parameter set $(\vec X)$ two extra possible combinations of the surface
parameters, which span the uncertainty of the $\chi^2$ minimum. Specifically, we impose a condition 
on $\chi^2$ that the ratio of maximal and minimal probabilities is $P_{min}/P_{max} = 1/2$ where
$P_{max,min}$ $\propto$ $\exp^{-\chi^2_{min,max}/2}$. The condition is then 
\begin{equation}\label{eq:condition1}
 \chi^2_{max} = 2ln(2) + \chi^2_{min},
\end{equation}
given by the dashed line in the Figure \ref{fig:CfinDfin}. The crossings between the
$\chi^2_{max}$ and the $C_{fin}$ and $D_{fin}$ lines gives the minimum and maximum values
which can be adopted for the surface parameters. This procedure gives a more realistic 
estimation of the uncertainty associated to the surface parameters, with respect to 
the rough and extreme condition of Section \ref{sec:Sensitivity} that the dynamical spinodal
must be embedded into the thermodynamical one. Still, this $\chi^2$ criterium implies that 
$C_{fin}$ and $D_{fin}$ are independent, which is certainly not the case.
The large range in $D_{fin} \pm \sigma$ can be further reduced  by looking which $D_{fin}$ 
values actually represent physical solutions. We have found that the very negative values 
of $D_{fin}$ are not physical by imposing the condition that the combined effect of surface
and Coulomb terms in eq. (\ref{eq:Cmatrix}) should quench the thermodynamical instability 
(i.e. $k \rightarrow 0$), and not amplify it. This condition can be expressed as
\begin{equation}\label{eq:condition2}
 \lim_{k\rightarrow0} e_2 (\mathcal{C}_k) = 0^+
\end{equation}
where $e_2(\mathcal{C}_k)$ is the lowest eigenvalue of the reduced matrix 
\begin{equation}
 \mathcal{C}_k = k^2 \mathcal{C}_{surf} + \frac{\alpha}{k^2} \mathcal{C}_{Coul}.
\end{equation}
Since the electrons play negligible role in the finite size spinodal \cite{Ducoin2007}, 
we can impose this condition to the nuclear matter without electrons, that is
\begin{equation}
\mathcal{C}_k = \begin{pmatrix}
         (C_{fin} + D_{fin})k^2 &  (C_{fin} - D_{fin})k^2 & 0\\
         (C_{fin} - D_{fin})k^2 &  (C_{fin} + D_{fin})k^2 & \frac{\alpha}{k^2}. \\
                     \end{pmatrix}    
\end{equation}
The eigenvalues $e$ of this matrix are 
\begin{equation}
 2e = \tr{(\mathcal{C}_k)} \pm \sqrt{ \tr^2{(\mathcal{C}_k)} - 4 \det{(\mathcal{C}_k})}
\end{equation}
which, after applying the condition (\ref{eq:condition2}), restricts the minimal value of
$D_{fin}$ parameter to the value 
\begin{equation}
 D_{fin}^{(min)} = - C^{(opt)}_{fin}.
\end{equation}
Summarizing the discussion, each parameter set $(\vec X)=(\{P_\alpha\},m^\ast,\Delta m^\ast, b)$ 
of our prior distribution will be associated to three different parameter sets, namely
\begin{itemize}
 \item  $\vec X, C_{fin}=C^{(opt)}_{fin},  D_{fin}=D1=-C^{(opt)}_{fin}$,
 \item  $\vec X, C_{fin}=C^{(opt)}_{fin},  D_{fin}=D3= D^{opt}_{fin} + \sigma$,
 \item $\vec X, C_{fin}=C^{(opt)}_{fin},  D_{fin}=D2 = (D1 + D2) /2$.
\end{itemize}

%For the specific case of our reference model Set 1 and Set 2, the uncertainty is estimated as follows:
%range around the minimum, the values being
%\begin{itemize}
% \item Set 1: $C_{fin}^{Set 1} = 52 \pm 18$, $D_{fin}^{Set 1} = -30 \pm 680$
% \item Set 2: $C_{fin}^{Set 2} = 46 \pm 18$, $D_{fin}^{Set 2} = -70 \pm 680$
%\end{itemize}

The so-defined prior set has been run through the filter given by the low density ab-initio calculation 
(LD) and the mass reproduction filter eq.(\ref{eq:probalikely}). Our posterior set is constituted of 
5412 models. %\cite{Carreau} 
It is interesting to remark that the low density filter \cite{Drischler2016} is much more constraining
than the mass filter. Applying this latter after the LD filter does not give any significant changes in
the parameter expectation values and variances. Still the mass filter allows determining reliable values
of the two gradient couplings, $C_{fin}$ and $D_{fin}$ which enter in the spinodal determination.

\begin{table*}[t!]
 \centering
 \bgroup
\def\arraystretch{1.3}
 \caption{The prior and posterior average values and variances. 
%The posterior is defined by the  low density filter \cite{Drischler2016}, and the values are taken from [Carreau]. The mass filter  doesn't give any significant changes in the parameter ranges but adds the values of the two  additional parameters, $C_{fin}$ and $D_{fin}$.
}
 \label{tab:param}
 \begin{tabular}{cccccccccccccccccccc}
 \hline
 \hline
   & & & & $n_{sat}$ & $E_{sat}$ & $K_{sat}$ & $Q_{sat}$ & $Z_{sat}$ & $E_{sym}$ & $L_{sym}$ 
   & $K_{sym}$ & $Q_{sym}$ & $Z_{sym}$ & $\frac{m^{\ast}}{m}$ & $\frac{\Delta m^{\ast}}{m}$ & $b$ & $C_{fin}$ & $D_{fin}$ \\
   & & && $[\frac{1}{fm^{3}}]$  & $[MeV]$  & $[MeV]$  & $[MeV]$  & $[MeV]$ & $[MeV]$ & $[MeV]$ 
   & $[MeV]$ & $[MeV]$ & $[MeV]$ &  &  & $[MeV]$ & $[\frac{MeV}{fm^{5}}]$ & $[\frac{MeV}{fm^{5}}]$ \\
\hline
    Prior     & & Av & &$0.16$&$-16$&$230$&$0$   &$0$   &$32$&$45$&$-100$&$0$   &$0$   &$0.7$&$0.1$ &$5.5$& & \\
	& & $\sigma$ & &$0.01$&$1$  &$40$ &$1000$&$3000$&$6$ &$35$&$300$ &$2000$&$5000$&$0.1$&$0.1$ &$4.5$ & & \\
    
        %& max &$0.17$&$-15$&$270$&$1000$ &$3000$ &$38$&$80$&$200$ &$2000$ &$5000$ &$0.8$&$0.2$ &$10$& & \\
	%        & min &$0.15$&$-17$&$190$&$-1000$&$-3000$&$26$&$10$&$-400$&$-2000$&$-5000$&$0.6$&$0.1$ &$1$ & & \\
	        \hline
    LD %\cite{Carreau} 
    & & Av & &$0.164$&$-15.29$&$234$&$-31$&$-146$&$30.7$&$43.7$&$-202$&$-253$&$-114$&$0.70$&$0.10$&$5.2$&$30$ & $115$  \\
    + mass filter & & $\sigma$ & & $0.005$&$0.25$&$23$ &$362$&$1728$&$0.8$&$3.7$&$42$&$673$&$2868$&$0.06$&$0.06$&$2.6$& $9$& $143$ \\
%     Mass   & max & & & & & & & & & & & & & & $30$ & $115$ \\
%     filter & min & & & & & & & & & & & & & & $9$& $143$\\
 \hline
 \hline
 \end{tabular} 
 \egroup
\end{table*}

 \begin{figure}[b!]
    \centering
    %%% == trim={<left> <lower> <right> <upper>}
    %\includegraphics[scale=0.28]{./Figure6.eps}
    \includegraphics[trim=0cm 0cm 0cm 0.1cm,clip=true,scale=0.28]{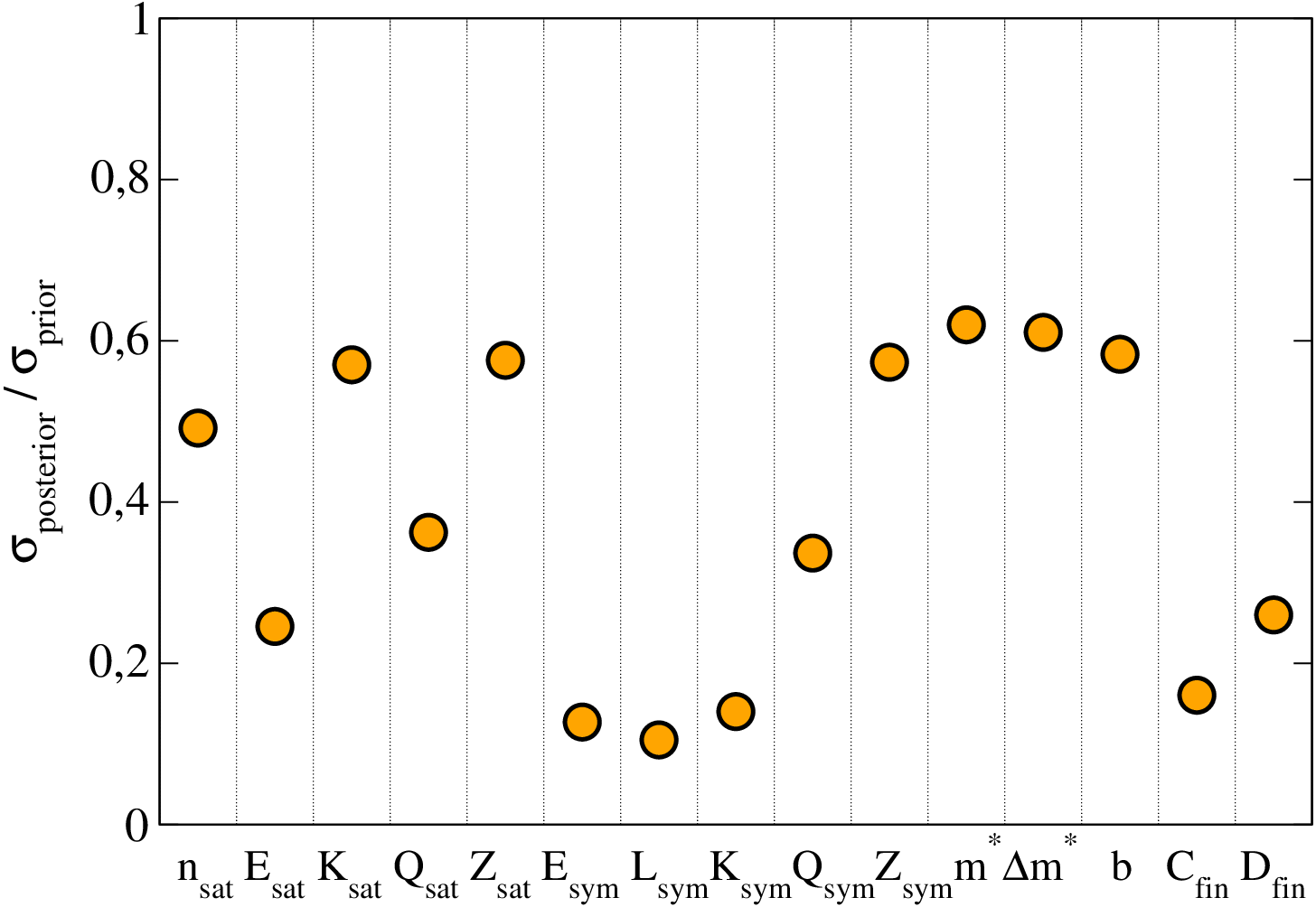}
    \caption{The representation of the parameter restrictions imposed by LD and mass filter
    and compared to the initially defined prior.}
%    while the $C_{fin}$ and $D_{fin}$ ranges are defined through the applied mass filter.}
    \label{fig:prior}
\end{figure}

Figure \ref{fig:prior} presents the variance reduction after 
the LD and mass filter in each of the parameters, as given in 
table \ref{tab:param}. It is represented as the ratio of the posterior and
prior variances for each of the model parameters. In this representation, the 
parameters with small value of the ratio are significantly constrained through
the LD filtering method compared to the prior range. As seen in Figure 
\ref{fig:prior}, the filtering procedure is very constraining for all parameters 
except the fourth order derivatives ($Z_{sat}$ and $Z_{sym}$), which in any case 
are not strongly influential in the determination of the CC point (see Section 
\ref{sec:Sensitivity}). The reduced impact on quantities like $K_{sat}$, $n_{sat}$ 
and $m^\ast$ can be understood from the fact that our prior distribution from table 
\ref{tab:param} takes into account the actual present knowledge of the different 
parameters from nuclear physics experiments. The quantities $K_{sat}$, $n_{sat}$ and
$m^\ast$ are already very well constrained, meaning that their prior dispersion is 
already very small, which explains the reduced effect of the filter. 

\begin{figure}[b!]
    \centering
    \includegraphics[scale=0.28]{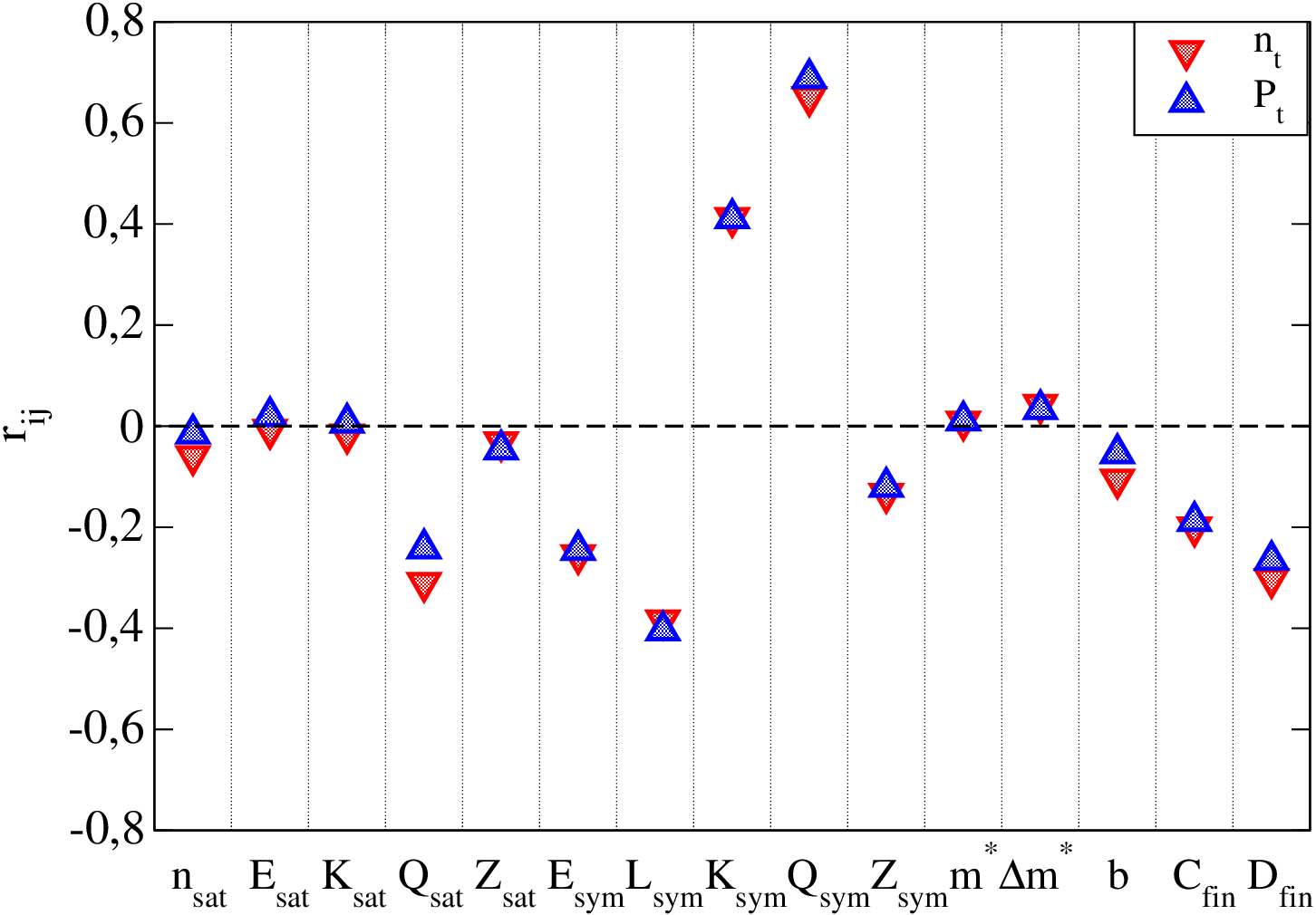}
    \caption{The correlation indices values $r_{ij}$ between the model parameters ($i$) and
    the CC transition point density $n_t$ and pressure $P_t$ ($j$).  }
    \label{fig:rij}
\end{figure}

%the effect is most pronounced for the parameters for which a wide range of values was 
%considered to start with in the prior set, such as $L_{sym}$ and $K_{sym}$. 
Finally, the question we want to address  is how much the CC point is correlated to the 
model parameters.
%Since the MM method has the possibility to span the full parameter space we are 
%calculating the correlations between the $n_t$ and $P_t$, and the model parameters for 
%the 1804 MMs that pass the LD  filter tripled by the three possible $C_{fin}-D_{fin}$
%combinations for each of them. 
The values of 
correlation indexes $r_{ij}$, where $i = n_t$, $P_t$ and $j = \{P_{\alpha}\}$, $m^{\ast}/m$, 
$\Delta m^{\ast}/m$, $b$, $C_{fin}$, $D_{fin}$, are given in Figure \ref{fig:rij}. The well determined
LO parameters have, as usually found, very little influence on the CC point. However, along with the
expected strong correlation with the $L_{sym}$ parameter, the $n_t$ and $P_t$ show also a
significant correlation with  the isovector curvature  $K_{sym}$ as well as with the $N=3$ order parameters, 
especially with $Q_{sym}$. This is in good qualitative agreement with the findings of the sensitivity 
analysis in Section \ref{sec:Sensitivity}. We can also see that the important role on the transition point
played by the behavior at very low density, as expressed by the influence of the $b$ parameter 
in Section \ref{sec:Sensitivity}, is strongly reduced in the correlation matrix. This can be understood 
from the important constraint at very low density given by the EFT calculations (LD filter).

\section{Conclusion}\label{sec:Conclusion}

The influence of the uncertainties in the EOS parameters on the crust-core
phase transition (CCPT) point was explored within a meta-modeling (MM)
approach. We have shown that at such low densities, the MM approach has to go 
beyond the second order expansion of energy per particle and take the higher orders 
into account in order to accurately reproduce the spinodal shapes of reference
models. This underlines the important influence of high order empirical parameters 
beyond the symmetry energy at saturation $E_{sym}$ and its slope $L_{sym}$, on the 
properties of the transition. This is evident from the obtained high correlation index
between the CC density and pressure values and the $N=2,3$ isovector empirical EOS 
parameters, as in Figure \ref{fig:rij}.  

A quantitative estimation of the model dependence on the transition point was obtained 
through a complete Bayesian analysis of the EoS parameters, where the multi-dimensional 
EoS parameter space has been constrained to reproduce the low density N3LO effective 
field theory predictions of Ref. \cite{Drischler2016}, and the gradient couplings have been 
constrained to reproduce experimental nuclear masses.

Our final predictions for the CC  transition density and pressure are 
$n_t = (0.071 \pm 0.011)$ $fm^{-3}$ and $P_t = (0.294 \pm 0.102)$ $MeV fm^{-3}$,
respectively. 

It might be disappointing to remark that the obtained uncertainty on the transition point 
is not much different from the present dispersion coming from compilations of phenomenological relativistic  and non-relativistic models (see for instance \cite{Ducoin:2011fy}), 
in spite of the fact that all used MMs satisfy the
ab-initio constraints, which are very restrictive at low density. 
The reason for this is possibly the fact that the reduced interval of $L_{sym}$ implied 
by the ab-initio filter with respect to the range of popular models, is compensated by 
wider uncertainty in the isovector high order and surface parameters. Improved predictions 
thus will require further experimental constraints for very low density asymmetric matter, 
such as for e.g. measurement of neutron skins \cite{Boso2018}. \\%Debi: Add references for upcoming
%experiments on neutron skins? \cite{?}

\section*{Acknowledgements}
%The authors thank Thomas Carreau for providing the table of MMs passing the low density 
%filter and the code for the mass filtering on which the method to extract the $C_{fin}$  and $D_{fin}$ parameters was built. 
Discussions with J\'er\^ome Margueron are gratefully acknowledged. S.A. acknowledges the PHAROS COST 
Action (CA16214) and ``NewCompstar`` COST Action (MP1304) for partial support of this project 
through the Short term scientific mission (STSM), and CNRS/In2p3 (Master project MAC) for partial support.

\bibliographystyle{plain}
\bibliography{SAntic_arXvs1}

\end{document}